\documentclass[a4paper,reprint,twocolumn,superscriptaddress,showkeys,aps,pre]{revtex4-1}
\usepackage{graphicx}
\usepackage{latexsym}
\usepackage{amsmath}
\usepackage{amssymb}
\usepackage{enumitem}
\usepackage{titlesec}
\usepackage{relsize}
\usepackage[usenames]{color}

\titleformat{\section}{\normalfont\normalsize\filright\bfseries}{\thesection}{}{}
\titlespacing\section{0pt}{12pt plus 4pt minus 2pt}{4pt plus 2pt minus 2pt}

\titleformat{\subsection}[runin]{\normalfont\normalsize\filright\bfseries}{(\thesubsection)}{}{}
\titlespacing\subsection{0pt}{12pt plus 4pt minus 2pt}{4pt plus 2pt minus 2pt}

\makeatletter
\def\blfootnote{\xdef\@thefnmark{}\@footnotetext}
\makeatother

\renewcommand{\figurename}{Fig.} 
\def\fnum@figure{\figurename\nobreakspace\textbf{\thefigure}}

\newenvironment{Abstract}{\bf\noindent}{\normalfont}

\newenvironment{significancetext}{\section*{Significance}\it\noindent\small}{\normalfont}

\begin{document}

\title{Stress controls the mechanics of collagen networks}

\author{Albert James Licup}\thanks{AJL, SM, and AS are co-first authors.}
\affiliation{Department of Physics and Astronomy, VU University Amsterdam, 1081HV The Netherlands}
\author{Stefan M\"unster}\thanks{AJL, SM, and AS are co-first authors.}
\affiliation{Center for Medical Physics and Technology, Biophysics Group, Friedrich-Alexander Universit\"at Erlangen-N\"urnberg, 91052 Erlangen, Germany}
\author{Abhinav Sharma}\thanks{AJL, SM, and AS are co-first authors.}
\affiliation{Department of Physics and Astronomy, VU University Amsterdam, 1081HV The Netherlands}
\author{Michael Sheinman}
\affiliation{Department of Physics and Astronomy, VU University Amsterdam, 1081HV The Netherlands}
\author{Louise M. Jawerth}
\affiliation{Department of Physics, Harvard University, Cambridge, MA 02138}
\author{Ben Fabry}
\affiliation{Center for Medical Physics and Technology, Biophysics Group, Friedrich-Alexander Universit\"at Erlangen-N\"urnberg, 91052 Erlangen, Germany}
\author{David A. Weitz}
\affiliation{Department of Physics, Harvard University, Cambridge, MA 02138}
\affiliation{School of Engineering and Applied Sciences, Harvard University, Cambridge, MA 02138}
\author{Fred C. MacKintosh}\thanks{To whom correspondence should be addressed. Email: fcmack@gmail.com}
\affiliation{Department of Physics and Astronomy, VU University Amsterdam, 1081HV The Netherlands}
\maketitle

\blfootnote{AJL, AS, MS, and FCM conceived and developed the model and simulations. AJL and AS performed the simulations. SM, LMJ, BF and DAW designed the experiments. SM and LMJ performed the experiments. All authors contributed to the writing of the paper.}
\blfootnote{The authors declare no competing financial interests.}

\begin{Abstract}
Collagen is the main structural and load-bearing element of various connective tissues, where it forms the extracellular matrix that supports cells.
It has long been known that collagenous tissues exhibit a highly nonlinear stress-strain relationship~\cite{art:Fung1967,art:Humphrey}, although the origins of this nonlinearity remain unknown~\cite{art:McMahon}.
Here, we show that the nonlinear stiffening of reconstituted type I collagen networks is controlled by the applied stress, and that the network stiffness becomes surprisingly insensitive to network concentration. 
We demonstrate how a simple model for networks of elastic fibers can quantitatively account for the mechanics of reconstituted collagen networks.
Our model points to the important role of normal stresses in determining the nonlinear shear elastic response, which can explain the approximate exponential relationship between stress and strain reported for collagenous tissues~\cite{art:Fung1967}. This further suggests new principles for the design of synthetic fiber networks with collagen-like properties, as well as a mechanism for the control of the mechanics of such networks.
\end{Abstract}

\begin{significancetext}
We report nonlinear rheology experiments on collagen type I networks, which demonstrate a surprising concentration independence of the network stiffness in the nonlinear elastic regime. We develop a model that can account for this, as well as the classical observations of an approximate exponential stress-strain relationship in collagenous tissues, for which a microscopic model has been lacking. Our model also demonstrates the importance of normal stresses in controlling the nonlinear mechanics of fiber networks.
\end{significancetext}

\section*{Introduction}
Collagen type I is the most abundant protein in mammals where it serves as the primary component of many load-bearing tissues, including skin, ligaments, tendons and bone. Networks of collagen-type I fibers exhibit mechanical properties that are unmatched by man-made materials. A hallmark of collagen and collagenous tissues is a dramatic increase in stiffness when strained. Qualitatively, this property of strain stiffening is shared by many other biopolymers, including intracellular cytoskeletal networks of actin and intermediate filaments~\cite{art:Gardel,art:Storm,art:KroyBausch,art:LinPRL}. On closer inspection, however, collagen stands out from the rest: it has been shown that collagenous tissues exhibit a regime in which the stress is approximately exponential in the applied strain~\cite{art:Fung1967}. The origins of this nonlinearity are still not known~\cite{art:McMahon,art:Motte}, and existing models for biopolymer networks cannot account quantitatively for collagen. In particular, it is unknown if the nonlinear mechanical response of collagen originates at the level of the individual fibers~\cite{art:Gardel,art:Storm,art:Yang2009,art:MacKintosh} or arises from nonaffine network deformations as suggested by numerical simulations~\cite{art:Head2003PRE,art:Onck,art:Chandran,art:Heussinger2006,art:HeadPRL,art:Wilhelm,art:Piechocka1}.

Here, we present both experimental results on reconstituted collagen networks, as well as a model that quantitatively captures the observed nonlinear mechanics. Our model is a minimal one, of random networks of elastic fibers possessing only bending and stretching elasticity. This model can account for our striking experimental observation that the stiffness of collagen becomes \emph{independent of protein concentration} in the nonlinear elastic regime, over a range of concentrations and applied shear stress. Our model highlights the importance of local network geometry in determining the strain threshold for the onset of nonlinear mechanics, which can account for the concentration independence of this threshold that is observed for collagen~\cite{art:Motte,art:Piechocka1}, in strong contrast to other biopolymer networks. Finally, our model points to the important role of normal stresses in determining the nonlinear shear elastic response, including the approximate exponential relationship between stress and strain reported for collagenous tissues~\cite{art:Fung1967}.

\section*{Results and Discussion}

In contrast to most synthetic polymer materials, biopolymer gels are known to exhibit a strong stiffening response to applied shear stress, in some cases leading to a more than 100-fold increase in the shear modulus, at strains as low as 10\% or less, before network failure~\cite{art:Gardel,art:Storm,art:KroyBausch,art:LinPRL}. 
Here, we perform rheology experiments on reconstituted networks of collagen type I, a key component of many tissues. 
We measure the differential shear modulus $K=\partial\sigma/\partial\gamma$ relating the shear stress $\sigma$ to the strain $\gamma$. We plot this in Fig.~\ref{fig:experimentdetail}a as a function of the applied stress. 
At low stress (and strain), we observe a linear elastic response with $K=G$, the linear shear modulus, which increases with collagen concentration. 
These networks also exhibit a strong increase in their stiffness $K$ above a threshold stress that increases with concentration. 
Remarkably, for network concentrations ranging from 0.45 to 3.6 mg/ml, the modulus becomes \emph{insensitive to concentration} in the nonlinear regime, where $K$ increases approximately linearly with $\sigma$: here, for a given sample preparation (e.g., polymerization temperature), the various $K$ vs $\sigma$ curves overlap, in spite of the fact that the linear moduli of these samples vary by two orders of magnitude.

Moreover, the approximate linear dependence of $K$ on $\sigma$ in our reconstituted networks is consistent with the empirically established exponential dependence of stress on strain in collagenous tissues~\cite{art:Fung1967}, 
since $\sigma\propto\exp\left(\gamma/\gamma_0\right)$ implies that $K=d\sigma/d\gamma\propto\sigma$.
While qualitatively similar stiffening with applied stress has been reported for other biopolymers~\cite{art:Gardel,art:KroyBausch,art:LinPRL,art:Muenster,art:BroederszRMP}, both the linear dependence of $K$ on $\sigma$ and the insensitivity of the nonlinear stiffening to network concentration appear to be unique to collagen.

\subsection*{Physical picture.}
\begin{figure*}[p!]
\centering
\includegraphics[width=0.9\textwidth]{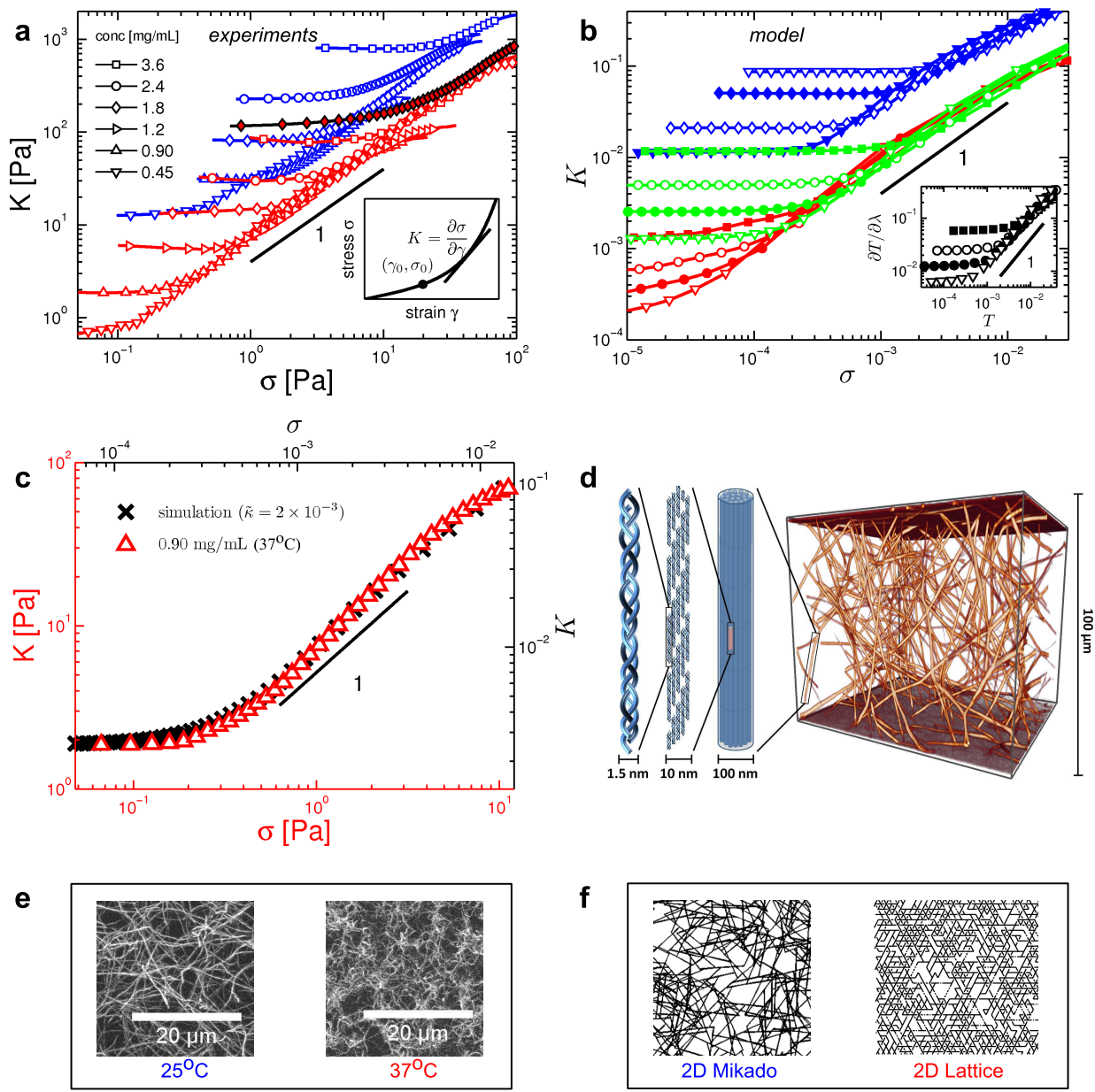}
\caption{Stiffening of reconstituted collagen type I networks. 
(a) Differential shear modulus $K$ vs shear stress $\sigma$ for reconstituted collagen type I networks at varying protein concentrations and different polymerization temperatures (red: $37\,^{\circ}\mathrm{C}$, blue: $25\,^{\circ}\mathrm{C}$). Lines of unit slope serve as visual guides. The filled red diamonds represent a network polymerized at $37\,^{\circ}\mathrm{C}$ at a concentration of 1.8 mg/ml with $0.2\%$ glutaraldehyde cross-linkers. The inset is a schematic of a typical stress vs strain curve indicating the stiffness $K$ as the tangent or differential shear modulus and the point $(\gamma_0,\sigma_0)$ at the onset of stiffening. (b) Simulation results for 2D (red) and 3D (green) lattice-based and 2D Mikado (blue) networks for various reduced bending rigidities $\tilde\kappa=10^{-2}$ ($\mathsmaller{\blacksquare}$), $\tilde\kappa=4\times 10^{-3}$ ($\mathlarger{\mathlarger{\mathlarger{\circ}}}$), $\tilde\kappa=2\times 10^{-3}$ ($\mathlarger{\mathlarger{\mathlarger{\bullet}}}$), $\tilde\kappa=10^{-3}$ ($\triangledown$), $\tilde\kappa=6\times 10^{-4}$ ($\blacklozenge$), $\tilde\kappa=2\times 10^{-4}$ ($\lozenge$), and $\tilde\kappa=10^{-4}$ ($\blacktriangledown$). The lattice-based networks (red and green) have connectivity $z=3.2$ corresponding to an aspect ratio $L/\ell_0=5$, while the Mikado networks (blue) have $z=3.6$ with $L/\ell_0=11$. We see that changing $z$ affects the overall magnitude of the moduli, but not the functional form of the stiffening response.
The inset shows stiffness vs stress curves from a 3D lattice-based network simulation under volume-preserving extension, where $T$ is the extensional stress and $\lambda$ is the extension ratio. (c) Comparison of $K$ vs $\sigma$ curves obtained from experiment ($\vartriangle$) and 3D lattice-based network simulation ($\times$) under shear. Multiplicative factors for the stiffness and stress axes have been chosen for coincidence of the linear modulus and the stress at the onset of nonlinearity. (d) A 3D confocal image of a reconstituted collagen type I network shows a highly branched local geometry (right). Collagen fibers are hierarchically assembled of fibrils (diameter: 10nm) which in turn consist of staggered collagen molecules (diameter: 1.5nm). The overall fiber diameter is of order 100nm, which makes the fibers sufficiently rigid enough to be modeled as an elastic beam. (e) Confocal images show differences in network geometry at different polymerization temperatures. Polymerizing collagen at $25\,^{\circ}\mathrm{C}$ creates networks of straighter, less branched fibers in contrast to networks polymerized at $37\,^{\circ}\mathrm{C}$. (f) The 2D network geometries used in the simulations.}
\label{fig:experimentdetail}
\end{figure*}

Although surprising at first sight, the features seen in Fig.\ \ref{fig:experimentdetail}a can be understood in simple physical terms for athermal networks of fibers that are soft to bending and where the nonlinear network response is controlled by stress. 
At low stress, if the elastic energy is dominated by soft bending modes, the linear shear modulus $G$ should be proportional to the fiber bending rigidity $\kappa$.
Of course, $G$ also depends on the density of collagen, as can be seen in Fig.\ \ref{fig:experimentdetail}a.
The concentration can be characterized in geometric terms by $\rho$, the total length of fiber per unit volume. 
Since $\kappa$ has units of energy$\times$length, while $G$ has units of energy per volume, we expect that $G\propto\kappa\rho^2$~\cite{art:Kroy,art:Satcher}. 
Since stress has the same units as $G$, similar arguments apply to the characteristic stress $\sigma_0\propto\kappa\rho^2$, above which the response becomes nonlinear. 
For $\kappa=0$, such networks become entirely \emph{floppy} and their rigidity depends on other stabilizing effects, or \emph{fields}, including applied stress~\cite{art:BroederszRMP,art:Alexander,art:Wyart,art:Broedersz2011NatPhys,art:Sheinman}. 
Thus, when the applied stress $\sigma$ becomes large enough to dominate the initial stability due to fiber bending resistance, it is expected that $K$ will increase proportional to $\sigma$, in a way analogous to the linear dependence of magnetization on field in a paramagnetic phase. 
Combining these observations, one obtains an approximate stiffening given by $K\propto G\times\left(\sigma/\sigma_0\right)^m$, where the slope $m=1$. 
Here, since $G$ and $\sigma_0$ have the same dependence on concentration, one obtains a nonlinear stiffness $K$ that becomes insensitive to concentration. 
Interestingly, this behavior is neither expected nor observed for F-actin and intermediate filament networks, which are not bend-dominated and exhibit a stronger nonlinear stiffening regime, in which $K\propto\sigma^{3/2}$~\cite{art:Gardel,art:LinPRL}.

\subsection*{Model.}
To test this simple physical picture, as well as uncover the mechanisms of collagen elasticity in more detail, we study simple/minimal computational models of fiber networks, specifically, 2D and 3D lattice-based networks~\cite{art:Broedersz2011SM,art:Broedersz2012PRL,art:Feng2014} and 2D Mikado networks~\cite{art:Head2003PRE,art:Wilhelm,art:Conti}. 
It is known that the mechanical stability and rigidity of networks depends on their connectivity, which can be characterized by the coordination number $z$, defined by the number of fiber segments meeting at a junction. 
Prior imaging of collagen networks~\cite{art:Lindstrom} report an average connectivity $z\simeq3.4$. 
Importantly, this places such networks well below the \emph{isostatic} or critical connectivity of $z=4$ in 2D or $z=6$ in 3D required for mechanical
stability of networks with only spring-like stretching energies~\cite{art:Maxwell}.
As a result, the linear elastic properties are expected to be governed by other energies, such as fiber bending~\cite{art:Head2003PRE,art:Onck,art:Wilhelm,art:BroederszRMP,art:Kroy,art:Satcher}, as well as by the distance of $z$ from its critical value~\cite{art:Alexander,art:Sheinman}.  
Thus, we generate our networks within a range of $z$, straddling the experimentally relevant values. 
Specifically, our 2D and 3D lattice-based networks are created with $z=3.2$ and our Mikado networks have $z=3.6$.
As we show below, properties such as the linear modulus $G$ and the strain threshold for the onset of nonlinear elasticity depend on $z$, although the overall form of the nonlinear regime is unaffected.

In our model, as in our experiments, we impose a volume-preserving simple shear strain $\gamma$ and minimize the total elastic energy $\mathcal{H}$ of the network, consisting of the sum of elastic energies of the individual fibers. The elastic energy of a fiber is calculated using a discrete form of the extensible wormlike-chain model that accounts for both local stretching and bending~\cite{art:Conti} (also Supporting Information). The network stiffness $K$ is calculated as
\begin{equation}
K=\frac{1}{V}\frac{\partial^2\mathcal{H}}{\partial\gamma^2},
\label{eqn:G}
\end{equation}
where $V$ is the volume of the system. Since $K$ depends on the energy per unit volume, and the energy involves an integral along the contour of all fibers in the system, $K$ is naturally proportional to the total length of fiber per volume, $\rho$, which is proportional to the protein concentration $c$. 
Thus, $K$ can be expressed as (Supporting Information)
\begin{equation}
K=\mu\rho\mathcal{K}\left(\gamma,\tilde\kappa\right),
\label{eqn:dimlessG}
\end{equation}
where $\mu$ is the fiber stretching modulus and $\tilde\kappa=\kappa/\mu\ell_0^2$ is a dimensionless measure of the relative bend-stretch stiffness, with $\ell_0$ a measure of the spacing between filaments. 
Here, $\rho\propto\ell_0^{1-d}$.
For lattice-based networks, we define $\ell_0$ to be the lattice spacing, while for Mikado networks we use the average distance between crosslinks.
The shear stress $\sigma$ can be expressed in a similar fashion as $\sigma=\mu\rho\Sigma\left(\gamma,\tilde\kappa\right)$.
For a given network structure, $\mathcal{K}$ and ${\Sigma}$ are dimensionless functions of only $\gamma$ and $\tilde\kappa$. 

In our simulations, we determine both $K$ and $\sigma$ for various values of $\kappa$.
We do this for networks with $\mu=1$ and $\ell_0=1$.
Thus, our simulation values of both moduli and stress are in units of $\mu/\ell_0^{d-1}$ in $d$ dimensions. 
We plot ${K}$ vs $\sigma$ in Fig.~\ref{fig:experimentdetail}b. For an elastic rod of diameter $2a$ and Young's modulus $E$, the parameter $\tilde\kappa$ is proportional to the fiber volume fraction $\phi$, since $\kappa=\pi a^4 E/4$, $\mu=\pi a^2 E$ and $\tilde\kappa=a^2/(4\ell_0^2)\propto\phi$~\cite{art:Head2003PRE,art:Wilhelm,art:Conti}. We thus consider values around $\tilde\kappa\lesssim 10^{-3}$ to compare with experiments, where the protein volume fraction varies over a range of approximately one decade around 0.1\%. 

Consistent with our experiments, our model networks also show an approximately linear relationship between stiffness $K$ and shear stress $\sigma$, as shown in Fig.~\ref{fig:experimentdetail}b~\cite{art:Broedersz2011SM}. 
We also study networks under extension, for which our model predicts a linear relationship between the stiffness and extensional stress, as shown in the inset to Fig.~\ref{fig:experimentdetail}b. Thus, our model can also account for prior experiments on collagenous tissues, which report such a linear relationship~\cite{art:Fung1967}. 
Moreover, both experiments and theory show a very surprising result in the \emph{stiffening regime}, 
where the $K$ vs $\sigma$ curves for different networks are seen to cluster around a \emph{common} line, and where networks of varying protein concentrations exhibit the \emph{same} stiffness at a given level of applied shear stress; i.e., the network stiffness $K$ becomes independent of network concentration and appears to be governed only by the applied stress in the nonlinear regime.

For low stress, the linear regime is indicated by a constant stiffness $K=G$, for which our model predicts the linear dependence on $\tilde\kappa$: $G\propto\rho\tilde\kappa\propto\kappa\rho^2$. This is consistent with both our observed increase of $G$ with collagen concentration in the experiments (Fig.~\ref{fig:conc}a), as well as with prior reports showing an approximate quadratic dependence of $G$ on concentration~\cite{art:Motte,art:Piechocka1}.
Moreover, to test whether for a given concentration $G$ increases with $\kappa$, we show data with glutaraldehyde (GA) cross-linkers, which increases the bending rigidity of collagen fibers~\cite{art:OldeDamink} (Fig.~\ref{fig:experimentdetail}a). Not only are these results consistent with the predicted increase in $G$, but the $K$ vs $\sigma$ curve still collapse onto the corresponding data for non-GA cross-linked networks in the stiffening regime. Thus, our model can account for the features observed in the experiments. For a more direct comparison we plot theoretical and experimental stiffening curves together in Fig.~\ref{fig:experimentdetail}c.
Moreover, both 2D and 3D results exhibit similar behavior, suggesting that stiffening is independent of dimensionality for a given local network geometry (Fig.~\ref{fig:experimentdetail}b).
\begin{figure}[ht]
\centering
\includegraphics[width=0.42\textwidth]{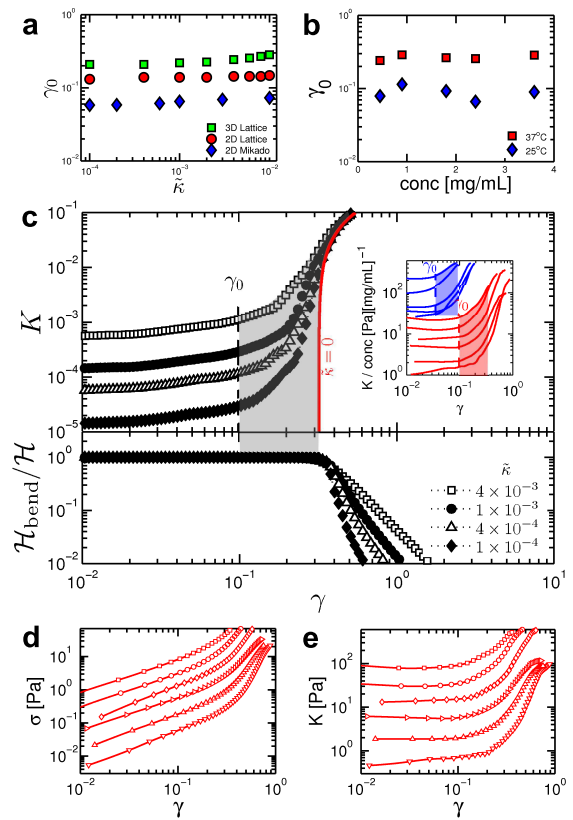}
\caption{Independence of characteristic strain $\gamma$ at the onset of stiffening on concentration. (Color online) (a) Onset strain $\gamma_0$ obtained from simulations vs fiber bending rigidity $\tilde\kappa$. (b) Experiments showing independence of $\gamma_0$ on protein concentration. (c) The upper panel shows the stiffness vs strain in a 2D lattice over the broad range of $\tilde\kappa$, and reveals a strain-stiffening regime highlighted by the shaded region. The inset shows the same data from Fig.~\ref{fig:experimentdetail}a normalized by concentration and plotted vs strain, with the dashed lines corresponding to the $\gamma_0$ values and the shaded regions highlighting the stiffening regimes. In both simulation and experiment, $\gamma_0$ is estimated as the strain at which the stiffness is roughly twice the linear modulus. The lower panel shows the overall contribution of bending energy to the total elastic energy in the network. (d,e) Experimental data for $37\,^{\circ}\mathrm{C}$ showing the strain dependence of the raw stress and stiffness. The symbols denote the same concentrations as shown in Fig.~\ref{fig:experimentdetail}a.}
\label{fig:theorydetail}
\end{figure}

In the nonlinear regime, the observed independence of $K/\sigma$ on concentration, and therefore on the typical spacing $\ell_0$ between fibers, suggests that the stiffening should be understood purely in geometrical terms, and quantities such as the characteristic strain $\gamma_0$ at the onset of stiffening should be independent of sample parameters such as concentration and $\kappa$. Figure~\ref{fig:theorydetail}a shows that $\gamma_0$ is indeed independent of $\tilde\kappa$, and is thus independent of both $\rho$ and $\kappa$, throughout the range $\tilde\kappa\lesssim 10^{-3}$. 
The strain threshold $\gamma_0$ does, however, depend on the the connectivity, $z$, as well as the type of the network, i.e., whether lattice-based or Mikado.
As we show in the Supporting Information, in the strongly bending-dominated limit, our model predicts a simple scaling dependence of $\gamma_0\propto\left(\ell_0/L\right)^2$ on the \emph{aspect ratio} $L/\ell_0$, where $L$ is the average length of the fibers. In general, $\gamma_0$ decreases with increasing $L/\ell_0$ or $z$.  For a given network type, lattice-based or Mikado, this aspect ratio is an entirely equivalent measure of connectivity to $z$: there is a one-to-one relationship between these two quantities, which increase (decrease) together. By construction, our networks have local coordination numbers strictly less than 4, which also represents the physiological upper bound of two fibers crossing at a cross-link. As the aspect ratio $L/\ell_0\rightarrow\infty$, $z\rightarrow4$ corresponding to the limit of very long fibers cross-linked many times to each other. Conversely, as $z$ decreases toward 3 (a branched structrure) the aspect ratio decreases toward unity. 
Thus, stiffening in our model networks is controlled by geometry, specifically via the aspect ratio $L/\ell_0$ or equivalently, the coordination number $z$. 

Collagen is known to form branched network structures~\cite{art:Lindstrom,art:Arevalo} (see Fig.~\ref{fig:experimentdetail}d), whose pore size scales as $1/\sqrt{c}\enskip$~\cite{art:Lang}. 
Changing the concentration only changes the degree of branching while preserving the local geometry, including the aspect ratio; i.e., networks at different concentrations look alike, apart from an overall scale factor. The onset strain $\gamma_0$ is then predicted to be independent of concentration, and indeed we observe this experimentally (Fig.~\ref{fig:theorydetail}b). Although this is consistent with prior experiments on collagen~\cite{art:Motte,art:Piechocka1}, it is in strong contrast to reports for other biopolymer networks~\cite{art:Gardel,art:KroyBausch,art:BroederszRMP,art:Kasza}.

\subsection*{Role of local network geometry.}
We can now understand quantitatively the features in our experiments based on three key assumptions: (i) the networks are athermal, (ii) are bend-dominated and (iii) their geometry at different concentrations is self-similar, i.e., the network structures at different concentrations are scale-invariant in that they are characterized by the same (aspect) ratio $L/\ell_0$. We test the last hypothesis by preparing collagen networks with different geometries. The structure of collagen networks strongly depends on the polymerization conditions, such as pH, ionic strength or temperature~\cite{art:Yang2009,art:Roeder,art:Achilli,art:Raub2007,art:Raub2008} (see Fig.~\ref{fig:experimentdetail}e).
Changing the local geometry, and specifically $L/\ell_0$ by changing the polymerization temperature does not affect the form of the stiffening response nor the collapse of the data in the nonlinear elastic regime, in either the model or the experiments, apart from a change in the ratio $K/\sigma$. The stiffening curves of networks with different geometries cluster around distinctly different curves of approximate unit slope (Fig.~\ref{fig:experimentdetail}a). Moreover, less branched networks show a lower $\gamma_0$ (Fig.~\ref{fig:theorydetail}b). This is consistent with simulation results when comparing Mikado with lattice-based networks (Fig.~\ref{fig:experimentdetail}b and Fig.~\ref{fig:theorydetail}a). To confirm that this is due to network geometry, and not to the temperature at which the rheology measurements are performed, we polymerize a network at $37\,^{\circ}\mathrm{C}$ and subsequently cool it to $25\,^{\circ}\mathrm{C}$. We then perform rheology measurements at $25\,^{\circ}\mathrm{C}$ and find that despite its larger linear modulus, the stiffening regime coincides with networks polymerized at $37\,^{\circ}\mathrm{C}$, demonstrating that network geometry, indeed, sets the prefactor $K/\sigma$ (Fig.~\ref{fig:conc}e).

To understand the stiffening mechanism, we first examine which of the two modes, stretching or bending, dominates the stiffening regime. Prior work has suggested that stiffening corresponds to a transition from bending- to stretching-dominated behavior~\cite{art:Onck}. Our simulations show that bending is dominant throughout the stiffening regime (Fig.~\ref{fig:theorydetail}c, Fig.~\ref{fig:energyratio}). When stretching modes finally become dominant, all $\mathcal{K}$ vs $\gamma$ curves converge, as shown in Fig.~\ref{fig:theorydetail}c. 
In most cases, this only occurs after the network stiffness has increased by more than an order of magnitude. Moreover, when stretching dominates, we find a distinct stiffening behavior characterized by $K\sim\sigma^{1/2}$ (Fig.~\ref{fig:rope})~\cite{art:Broedersz2011SM}. Thus, we find three distinct rheological regimes: (1) a linear elastic regime, (2) a bend-dominated stiffening regime, and (3) a stretch-dominated stiffening regime. Interestingly, the approximate $K\sim\sigma$ regime we observe in our collagen networks is consistent with the second of these regimes, which occurs before the transition from bend- to stretch-dominated behavior. 

The existence of a distinct bend-dominated nonlinear regime and the corresponding concentration-independent nonlinear response in Figs.~\ref{fig:experimentdetail}a and b depends crucially on the sub-isostatic nature of the networks, as well as on small values $\lesssim 10^{-2}$ of $\tilde\kappa$ in the model and volume fraction $\phi$ in experiments. The collagen networks we study here are, indeed, all sub-isostatic with respect to stretching alone~\cite{art:Broedersz2011NatPhys,art:Maxwell}, since $z\simeq3.4$ lies well below the critical connectivity of 6 in 3D (4 in 2D) at which pure spring networks first become stable. As either $\tilde\kappa$ or the aspect ratio $L/\ell_0$ increase, a transition to stretch-dominated linear elastic behavior is expected, even in 3D, where the networks remain clearly sub-isostatic~\cite{art:Broedersz2012PRL,art:BroederszRMP}. However, over the range $2.5\lesssim L/\ell_0\lesssim 11$ that we study here (Figs.\ \ref{fig:experimentdetail}b and S4), which includes reported collagen network structures, we consistently see effects of bend-dominated network response in our model, including the concentration-independent nonlinear behavior. Here, $\kappa$ acts as a stabilizing interaction or \emph{field} for networks in their linear elastic regime, with $G\propto\kappa$. The intermediate nonlinear regime, where we find $K\sim\sigma$ in our simulations and experiments, can be understood in terms of marginal stability together with the stabilizing effect of applied stress.
\begin{figure}[ht]
\centering
\includegraphics[width=0.47\textwidth]{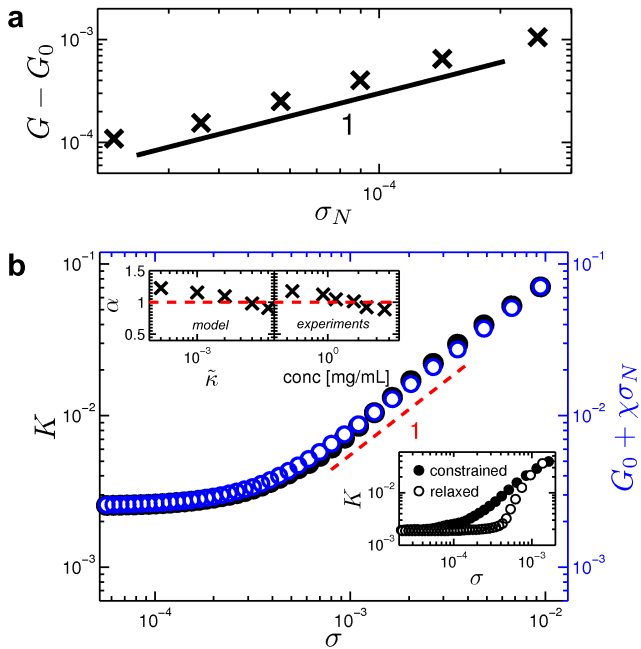}
\caption{Stiffening induced by normal stresses. (Color online) (a) The change in the linear shear modulus grows in direct proportion to an external normal stress $\sigma_N$ applied on the shear boundaries. (b) Comparison of $K$ (black) with $G_0+\chi\sigma_N$ (blue) vs shear stress in the linear and stiffening regimes. The local slope $\alpha=1$ (red dashed line) in the stiffening region is shown as a visual guide. The upper insets show the variation of $\alpha$ as a function of bending rigidity/protein concentration. The lower inset shows the reduction in network stiffness when the normal stress is removed.}
\label{fig:normal}
\end{figure}

\subsection*{Normal stresses.} 
Biopolymer networks, including collagen, have been shown to develop large negative normal stresses~\cite{art:Conti,art:Heussinger2007,art:Janmey}. This is in contrast to most elastic materials that exhibit positive normal stresses, as first demonstrated by Poynting~\cite{art:Poynting}, who showed elongation of wires under torsion. Biopolymer gels have been shown to do the opposite. In experiments, the constraint normal to the sample boundaries leads to the build-up of tensile stress
at these boundaries
when simple shear is imposed. These normal stresses can stabilize sub-marginal networks. In Fig.~\ref{fig:normal}a, we show that the linear modulus grows in direct proportion to an applied normal stress. We hypothesize that the network stiffness could arise from the normal stresses that develop under shear strain due to the imposed constraint at the boundaries:
\begin{equation}
K\simeq G_0 + \chi\sigma_N
\label{eqn:normal}
\end{equation}
where $G_0$ is the linear shear modulus in the absence of any normal stress $\sigma_N$ and $\chi$ is a constant. In Fig.~\ref{fig:normal}b, we show a direct comparison of $K$ and $G_0 + \chi\sigma_N$ vs $\sigma$, where $\sigma_N$ is independently measured in our simulations. The linear regime is characterized by $G_0$ in the absence of $\sigma_N$. In the stiffening regime, there is excellent agreement between $K$ and $G_0+\chi\sigma_N$, and both show the same local slope $\alpha\simeq 1$ consistent with the unit slope in Figs.~\ref{fig:experimentdetail}a and~\ref{fig:experimentdetail}b. Finally, we confirm our hypothesis by performing further relaxation of the networks when the normal stresses are released. In the lower inset of Fig.~\ref{fig:normal}b, by removing the normal stresses, we observe a significant reduction of the stiffness throughout the stiffening regime. This clearly supports the hypothesis that normal stresses control the stiffening of fiber networks under simple shear. Moreover, upon closer examination of the model predictions, we see a small but systematic evolution of the stiffening exponent $\alpha$ with $\tilde\kappa$, as shown in the upper left inset of Fig.~\ref{fig:normal}b. A similar evolution is seen in our experiments as a function of concentration, as shown in the right panel of the upper inset of Fig.~\ref{fig:normal}b. This agreement between experiment and model further justifies our identification of $\tilde\kappa$ with network concentration.

\section*{Concluding Remarks}
The development of normal stresses in these networks is intimately related to the volume-preserving nature of simple shear deformations, both in our rheology experiments and in our simulations. In our model, removal of the normal stress leads to a reduction in the volume of the system and a concomitant reduction in the stiffness. While collagenous tissues \emph{in vivo} are subject to more complex deformations, approximate volume conservation is valid in many cases, e.g., due to embedded cells~\cite{art:Fung1967}. 
Our results suggest that any volume-preserving deformation should lead to similar behavior in stiffness vs stress. 
In particular, in addition to accounting for the approximate exponential stress-strain relationship known empirically for collagen under extension~\cite{art:Fung1967}, our model also predicts that the nonlinear (Young's) modulus should become concentration independent for a given extensional stress, in a way similar to the case of simple shear. This can be seen in the inset to Fig.~\ref{fig:experimentdetail}b and Fig.~S7. 

The concentration-independence and collapse of the stress-stiffness curves seen in Fig.\ \ref{fig:experimentdetail}a appears to be unique to collagen networks, at least among biopolymers. Within our model, this property depends on three aspects: (1) the athermal and simple elastic response of the constituent fibers, (2) the bend-dominated response of the network in its linear elastic regime, and (3) the linear scaling of stiffness with stress, given by $K\sim\sigma^m$, where $m=1$. These properties are also expected for collagen type II, another fiber-forming type of collagen. For intracellular biopolymer networks, however, the latter property (3) is strongly violated: for actin and intermediate filament networks, a stronger stiffening, with $m\simeq3/2$ is observed. Interestingly, no such collapse or concentration-independence has been reported for those systems. 
One interesting consequence of the approximate collapse of the stress-stiffness relationship, combined with the lack of concentration dependence of strain (in Fig.\ \ref{fig:theorydetail}a,b), is that the local deformation of such matrices is expected to become nearly uniform in the nonlinear elastic regime, even in the presence of large local inhomogeneities in network density. This can have a stabilizing effect under excessive mechanical loading. The present work has identified the key properties that can form the basis for design of biomimetic networks with similar nonlinear properties to collagen. 

The importance of the nonlinear stiffness of collagen matrices comes in part from the inherent stability that such stiffening can impart to whole tissues: if collagen network elasticity were linear, then such networks would either fail or have to strain by more than 200-300\% under the maximum stresses in our experiments. Moreover, an initial soft elastic response of collagen also seems to be important physiologically: high stiffness due to excessive collagen production, e.g., during fibrosis, scar formation or around tumors is known to contribute strongly to pathological processes at the cellular level, where it can drive the so-called epithelial-to-mesenchymal transition and affect cell differentiation. Thus, both a soft initial linear response, as well as a strongly non-linear stiffening regime of collagen matrices are important for individual cells and tissues.
Apart from tissues with high content collagen, such as tendon and skin, most soft tissues have collagen content in the range of tenths of \% to a few \%, which corresponds to a range from our densest reconstituted networks up to about a decade higher in concentration~\cite{art:CollagenContent}. In such tissues, we expect nonlinear effects such as we report here to appear at shear stresses of order kPa, which is a level of stress easily reached, for instance, by traction forces of fibroblasts~\cite{art:Traction}. Thus, we expect the kind of stiffening reported here to be relevant to many soft tissues. 
While collagen networks have been known to exhibit nonlinear mechanics that is qualitatively similar to other biopolymer networks, it has become increasingly clear that the underlying mechanism of collagen stiffening differs from that of other biopolymers~\cite{art:Motte}. Not only does the present work shed light on the origins of collagen matrix mechanics, but it can also form a basis for the design of synthetic networks to mimic collagen and other extracellular matrices for tissue engineering.


\section*{Materials and Methods}
\subsection*{Polymerization of collagen networks.}
We dilute type I collagen (BD Biosciences, San Jose, CA) at 4$\,^{\circ}$C to the desired final concentrations between of 0.45 mg/ml and 3.6 mg/ml in 1x DMEM (Sigma Aldrich, St. Louis, MO) with 25 mM HEPES added and adjust the pH to 9.5 by addition of 1M NaOH. We fill the solution into the rheometer geometry preheated to 25$\,^{\circ}$C or 37$\,^{\circ}$C as indicated and allow for at least two hours of polymerization. To stiffen some samples, we pipette a solution of 0.2\% glutaraldehyde (Sigma) in 1x PBS (Lonza, Allendale, NJ) around the rheometer geometry once the networks have polymerized for 45 minutes and incubate these samples for three hours before performing experiments.%

\subsection*{Rheometry and data analysis.}
We perform the experiments on an AR-G2 rheometer (set to strain-controlled mode) or an ARES-G2 strain-controlled rheometer (both TA instruments, New Castle, DE) both fitted with a 25 mm PMMA disc as top plate and a 35 mm petri dish as bottom plate and set a gap of 400 $\mu$m. We prevent evaporation by sealing the samples with mineral oil, except for experiments on crosslinked collagen; here, we use a custom-built solvent trap, which allows for the addition of the crosslinking solution.
We monitor the polymerization of all samples by continuous oscillations with a strain amplitude of 0.005 at a frequency of 1 rad/s. Subsequently, we impose a strain ramp with a rate of 0.01/sec and measure the resulting stresses. We fit each stress-strain data set with a cubic spline interpolation and calculate its local derivative, which we then plot versus stress.%

\subsection*{Generation of disordered phantom networks.}
We take a $W\times W$ triangular lattice or a $W\times W\times W$ face-centered cubic (FCC) lattice of spacing $\ell_0$ to generate the disordered phantom network in two or three dimensions, respectively. In $d-$dimensions, the lattice occupies a volume $V=v_0 W^d$, where $v_0$ is the volume of a unit cell. Periodic boundaries are imposed to eliminate edge effects. A continuous chain of lattice bonds along a straight line forms a single fiber. The lattice vertices, having 6-/12-fold connectivity (i.e., coordination number) in 2D/3D, are freely-hinged cross-links, where fibers rotate about each other with no resistance. We reduce the average connectivity using the following procedure. In a 2D triangular lattice, we randomly select two out of the three fiber strands at a vertex on which we form a binary cross-link, i.e., with 4-fold connectivity. The remaining strand crosses this vertex as a phantom and does not interact with the other two strands. This is done at every vertex until all cross-links are binary. We further dilute the lattice by randomly removing bonds with probability $q=1-p$, where $p$ is the probability of an existing bond. After dilution, fibers that span the system size may still be present and these could lead to unphysical contributions on the macroscopic network stiffness. To avoid this, we make sure that every fiber has at least one diluted bond. All remaining dangling ends are further removed. Finally, nodes are introduced at the midpoint of every lattice bond so that the first bending mode on each bond is represented. The procedure just described effectively reduces both the average connectivity to $z<4$ and the average fiber length to $L=\ell_0/q$ and generates a disordered phantom triangular lattice~\cite{art:Broedersz2011SM}. A similar procedure as described can be implemented on the FCC lattice to generate a three-dimensional equivalent~\cite{art:Broedersz2012PRL}.

\subsection*{Generation of Mikado networks.}
We generate these networks~\cite{art:Conti} by random deposition of monodisperse fibers in the form of rods of length $L\ll W$ onto a two-dimensional $W\times W$ box, which occupies a volume $V=W^2$. Each rod's center of mass $(x_\mathrm{cm},y_\mathrm{cm})$ and orientation $\varphi$ relative to a fixed axis are each drawn from a uniform distribution. The box has periodic boundaries such that if a rod intersects any side of the box, it crosses over to the opposite side. A cross-link is assigned to the point wherever a given pair of rods intersect. Every time a rod is deposited, the cross-linking density $L/l_c$ is updated, where $l_c$ is the average distance between neighboring cross-links. Deposition stops as soon as the desired cross-linking density is achieved, after which all dangling ends are removed. Midpoint nodes are introduced on the rod between a pair of cross-links.

\subsection*{Discrete extensible wormlike chain model.}
The internal degrees of freedom  in the network is the set of spatial coordinates $\{r_i\}$ of all discrete nodes (i.e., cross-links, phantom nodes and midpoint nodes) on every fiber. Each fiber in the network is semiflexible, i.e., the elastic response to a given deformation is determined by both its stretching modulus $\mu$ and bending rigidity $\kappa$. When the network is deformed, the nodes undergo a displacement $\{r_i\}\rightarrow\{r'_i\}$. The extension of a fiber segment $\langle ij\rangle$ between nodes $i$ and $j$ along a fiber is given by $\delta\ell_{ij}=\ell'_{ij}-\ell_{ij}$, where $\ell'_{ij}=\|r'_j-r'_i\|$ and $\ell_{ij}=\|r_j-r_i\|$ is the rest length of the strand. Note that for lattice-based networks, $\ell_{ij}$ reduces to the bond rest length $\ell_0$ for all $\langle ij\rangle$. The total stretching energy of a fiber is then calculated by summing up the contributions of a chain of strands along its backbone:
\[\mathcal{H}_\mathrm{stretch}=\frac{1}{2}\mu\sum_{\langle ij\rangle}\ell_{ij}\left(\frac{\delta\ell_{ij}}{\ell_{ij}}\right)^2.\]
The bending of a fiber segment involves a triplet of consecutive nodes $\langle ijk\rangle$ along the backbone. The local curvature at node $j$ is estimated as $\|d\hat{t}/ds\|\approx\delta\hat{t}_j=\|\hat{t}_{jk}-\hat{t}_{ij}\|$, where $\hat{t}_{ij}$ is a unit vector oriented along $\langle ij\rangle$. The net contribution of consecutive segments $\langle ijk\rangle$ along a fiber leads to its bending energy
\[\mathcal{H}_\mathrm{bend}=\frac{1}{2}\kappa\sum_{\langle ijk\rangle}l'_j\left(\frac{\delta\hat{t}_j}{l'_j}\right)^2,\]
where $l'_j=\frac{1}{2}\left(\ell_{ij}+\ell_{jk}\right)$. Adding up $\mathcal{H}_\mathrm{stretch}+\mathcal{H}_\mathrm{bend}$ over all fibers in the network yields the total elastic energy.

\subsection*{Rheology simulation.}
We simulate rheology on the networks by imposing an affine simple shear strain $\gamma$. We fix the fiber stretching modulus $\mu=1$ and inter-filament spacing $\ell_0=1$. We vary $\kappa$ to probe a range of bending rigidities. We steadily increase the strain in $d\gamma$ steps to cover a strain range of $0.1\%$ to $1000\%$. At each strain step, the total elastic energy is minimized by relaxing the internal degrees of freedom using a conjugate gradient minimization technique~\cite{book:numrec}. Lees-Edwards boundary conditions are used when calculating the lengths of strands that cross the shear boundaries~\cite{art:LeesEdwards}. From the minimum energy $\mathcal{H}$, we extract the shear stress $\sigma$ and differential shear modulus $K$: \[\sigma=\tfrac{1}{V}\tfrac{\partial\mathcal{H}}{\partial\gamma},\enskip K\equiv\tfrac{\partial\sigma}{\partial\gamma}=\tfrac{1}{V}\tfrac{\partial^2\mathcal{H}}{\partial\gamma^2}.\]

\begin{acknowledgments}
AJL, AS, MS, and FCM were supported in part by FOM/NWO. SM and BF were supported by DFG. This work was supported in part by the National Science Foundation (DMR-1006546) and the Harvard Materials Research Science and Engineering Center (DMR-0820484).
\end{acknowledgments}

\makeatletter\def\@biblabel#1{#1.}\makeatother

\appendix


\renewcommand{\appendixname}{Supporting Information}

\renewcommand\thefigure{S\arabic{figure}}
\setcounter{figure}{0}

\renewcommand\theequation{S\arabic{equation}}
\setcounter{equation}{0}

\section*{Supporting Information}
\subsection*{Dimensionless Shear Modulus and Bending Rigidity}
For a homogeneous elastic rod~\cite{book:LandauLifshitz} of radius $a$ and Young's modulus $E$, the stretching modulus $\mu=\pi a^2E$ and bending rigidity $\kappa=\tfrac{\pi}{4}a^4E$. The bending length scale defined~\cite{art:Head2003PRE} as $\ell_b\equiv\sqrt{\kappa/\mu}$ is a length of order the rod diameter, since $\kappa/\mu\propto a^2$. So for every fiber segment of length $\ell_0$, we can express the bending rigidity in dimensionless form as $\tilde\kappa=\ell_b^2/\ell_0^2=\kappa/\mu\ell_0^2$. The differential shear modulus is derived from the energy density of the network, which requires calculating the total elastic energy per unit volume $\mathcal{U}=\frac{\mathcal{H}}{V}$. For a network occupying a volume $V$ and composed of $N$ fibers, this can be evaluated as a sum of the elastic energies of all semiflexible fibers $f$:
\[
\mathcal{U}=\frac{1}{V}\sum_{f=1}^{N}\left[\frac{\mu}{2}\int_f\left(\frac{d\ell(s)}{ds}\right)^2ds+\frac{\kappa}{2}\int_f\left|\frac{d\hat{t}(s)}{ds}\right|^2ds\right],
\]
where $d\ell(s)/ds$ and $|d\hat{t}(s)/ds|$ are the local length change and local curvature at a point $s$ on the fiber with unit tangent $\hat{t}(s)$. In a discrete network construction where the fibers are divided into a total of $n$ segments $\langle ij\rangle$, this can be approximated as
\begin{widetext}
\[
\begin{split}
\mathcal{U}&\simeq\frac{1}{V}\sum_{f=1}^{N} \left[\frac{\mu}{2}\sum_{\langle ij\rangle\in f}\ell_0\left(\frac{\delta\ell_{ij}}{\ell_0}\right)^2+\frac{\kappa}{2}\sum_{\langle ijk\rangle\in f}\ell_0 \left\|\frac{\hat{t}_{jk}-\hat{t}_{ij}}{\ell_0}\right\|^2\right]\\
&=\mu\frac{\ell_0}{V}\sum_{f=1}^{N} \left[\frac{1}{2}\sum_{\langle ij\rangle\in f}\left(\frac{\delta\ell_{ij}}{\ell_0}\right)^2+\frac{\tilde{\kappa}}{2}\sum_{\langle ijk\rangle\in f}\|\hat{t}_{jk}-\hat{t}_{ij}\|^2\right]\\
&=\mu\frac{N\ell_0}{V}\langle\mathcal{E}\left(\gamma,\tilde\kappa\right)\rangle_\mathrm{f}\\
&=\mu\frac{n\ell_0}{V}\langle\mathcal{E}\left(\gamma,\tilde\kappa\right)\rangle_\mathrm{s}.
\end{split}
\]
\end{widetext}
The quantity $\langle\mathcal{E}\left(\gamma,\tilde\kappa\right)\rangle_\mathrm{f}$ is a dimensionless elastic energy averaged over all fibers in the network, and $\langle\mathcal{E}\left(\gamma,\tilde\kappa\right)\rangle_\mathrm{s}$ is averaged over all fiber segments, of which there are $n$. Thus, since $n\ell_0$ is the total length of fiber in the system, 
\[
\mathcal{U}\approx\mu\rho\langle\mathcal{E}\left(\gamma,\tilde\kappa\right)\rangle_\mathrm{s},
\]
where $\rho$ is the network concentration in total fiber length per volume. 
Differentiating with respect to $\gamma$ yields the shear stress $\sigma=\mu\rho\Sigma\left(\gamma,\tilde\kappa\right)$, where the dimensionless stress
\[
\Sigma\left(\gamma,\tilde\kappa\right)=\frac{\partial}{\partial\gamma}\langle\mathcal{E}\left(\gamma,\tilde\kappa\right)\rangle_\mathrm{s}.
\]
Similarly, the dimensionless shear modulus $\mathcal{K}=\tfrac{\partial\Sigma}{\partial\gamma}$ is related to the shear modulus by $K=\mu\rho\mathcal{K}\left(\gamma,\tilde\kappa\right)$.

The concentration $\rho$ is also related to the fiber rigidity $\tilde\kappa$. For any given network structure of stiff rods, a segment of length $\ell_0$ and cross-section $a^2$ occupies a volume fraction $\phi\propto a^2\ell_0/\xi^3$, where the typical mesh size $\xi\sim\ell_0$. It follows that the concentration of fiber material $\rho\sim\phi\propto a^2/\ell_0^2$, and since the fiber rigidity $\tilde\kappa=\kappa/\mu\ell_0^2\sim a^2/\ell_0^2$, we obtain $\tilde\kappa\propto\rho$.

\begin{figure}[hb]
\centering
\centerline{\includegraphics[width=0.5\textwidth]{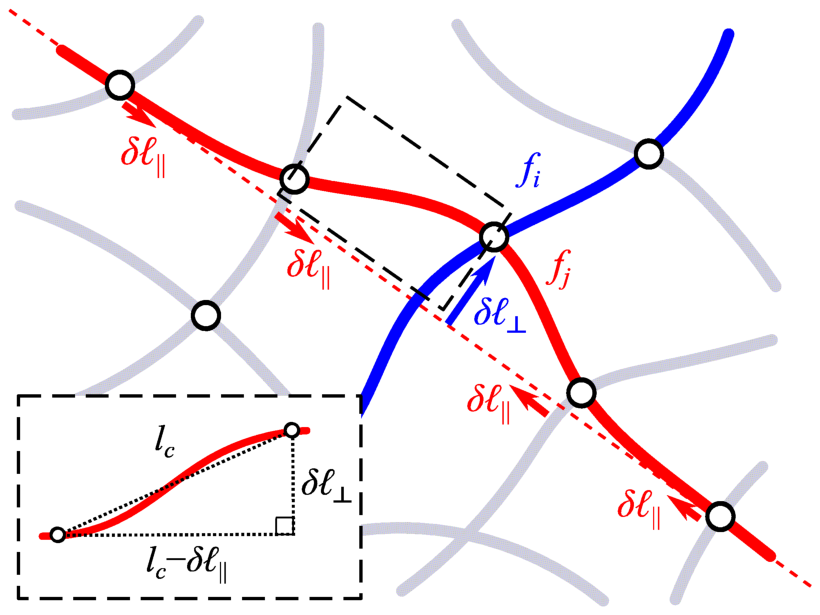}}
\caption{(Color online) Schematic showing two interacting fiber strands $f_i$ (blue) and $f_j$ (red) as well as other strands (gray). Circles denote points of mechanical constraints in the form of cross-links or branch points. Relaxation of $f_i$ along its backbone tugs $f_j$ inducing a transverse displacement $\delta\ell_\perp$ (blue arrow) and longitudinal displacements $\delta\ell_\parallel$ (red arrows). In a similar manner, $f_i$ experiences both displacements from its interaction with other strands. The zoom-in shows a simple first approximation geometric relation between these displacements.}
\label{fig:schematic}
\end{figure}

\begin{figure}[hb]
\centering
\centerline{\includegraphics[width=0.5\textwidth]{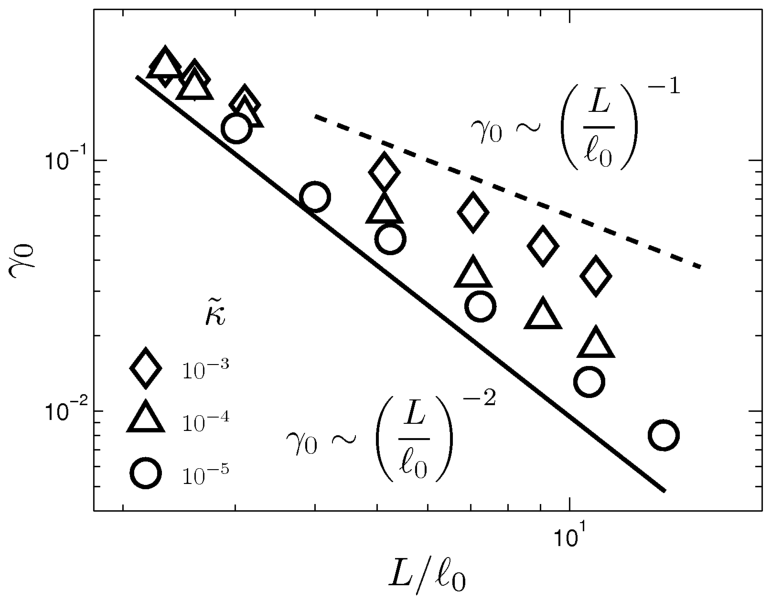}}
\caption{Critical strain $\gamma_0$ as a function of $L/\ell_0$. Networks with bend-dominated linear elasticity show a shift in the critical strain with the aspect ratio $L/\ell_0$ described by a line of best fit that agrees well with the model. For larger aspect ratios, the scaling crosses over to a weaker dependence.}
\label{fig:onset}
\end{figure}

\subsection*{Geometric Dependence of the Critical Strain}
The schematic in Fig.~\ref{fig:schematic} shows two fiber strands $f_i$ and $f_j$, each of length $\ell_0$ intersecting at a cross-link. The average length of the fibers in the network is $L$. Each fiber undergoes a backbone relaxation $\gamma L$ and we assume that the linear elastic response of the network is dominated by fiber bending interactions. The backbone relaxation of $f_i$ induces on $f_j$ a transverse displacement $\delta\ell_\perp\propto\gamma L$ and a longitudinal displacement $\delta\ell_\parallel$, as shown on the schematic. The longitudinal displacement which is a local retraction of $f_j$, is related to the transverse displacement as
\[
\begin{split}
\left(\ell_0-\delta\ell_\parallel\right)^2+\delta\ell_\perp^2&=\ell_0^2\\
\delta\ell_\parallel&=\ell_0-\sqrt{\ell_0^2-\delta\ell_\perp^2}\\
&=\frac{\delta\ell_\perp^2}{2\ell_0}+\frac{\delta\ell_\perp^4}{8\ell_0^3}+\cdots\\
\delta\ell_\parallel&\approx\frac{\delta\ell_\perp^2}{\ell_0},
\end{split}
\]
such that for small strains, we have $\delta\ell_\parallel\approx\frac{\gamma^2 L^2}{\ell_0}$. Since on average there are $L/\ell_0$ fibers attached to any given fiber, the total longitudinal displacement $\delta L_\parallel$ resulting from the backbone relaxations of these other fibers can be expressed as $\delta L_\parallel=\left(\frac{L}{\ell_0}\right)\delta\ell_\parallel\approx\frac{\gamma^2 L^3}{\ell_0^2}$. In the low strain limit $\delta L_\parallel\ll\gamma L$, i.e., the total longitudinal displacement of the fibers is negligible in comparison to their own backbone relaxations. The critical strain $\gamma_0$ is obtained when $\delta L_\parallel\approx\gamma L$, i.e.,
\begin{equation}
\label{eqn:gamma0}
\gamma=\gamma_0\sim\left(\frac{\ell_0}{L}\right)^2
\end{equation}
which corresponds to the onset of stiffening. Thus, onset $\gamma_0$ of stiffening is set by the geometric length scale aspect ratio $L/\ell_0$.

We emphasize that Eqn.~\ref{eqn:gamma0} applies to the asymptotic bend-dominated limit. This we observe in Fig.~\ref{fig:onset} for $L/\ell_0\lesssim 5$, which is the relevant parameter range in our comparison of simulation and experiment. The scaling crosses over to a weaker dependence for much larger $L/\ell_0$, where it appears to show $L^{-1}$ dependence. Such scaling has been reported in previous work~\cite{art:Broedersz2012PRL}. However, in contrast to the geometric mechanism presented in the current work, their $L^{-1}$ dependence is based on an energetic crossover from bending to stretching regimes. Whether the scaling changes from $L^{-2}$ to $L^{-1}$ needs to be further investigated.

\begin{figure*}[b]
\centering
\centerline{\includegraphics[width=0.9\textwidth]{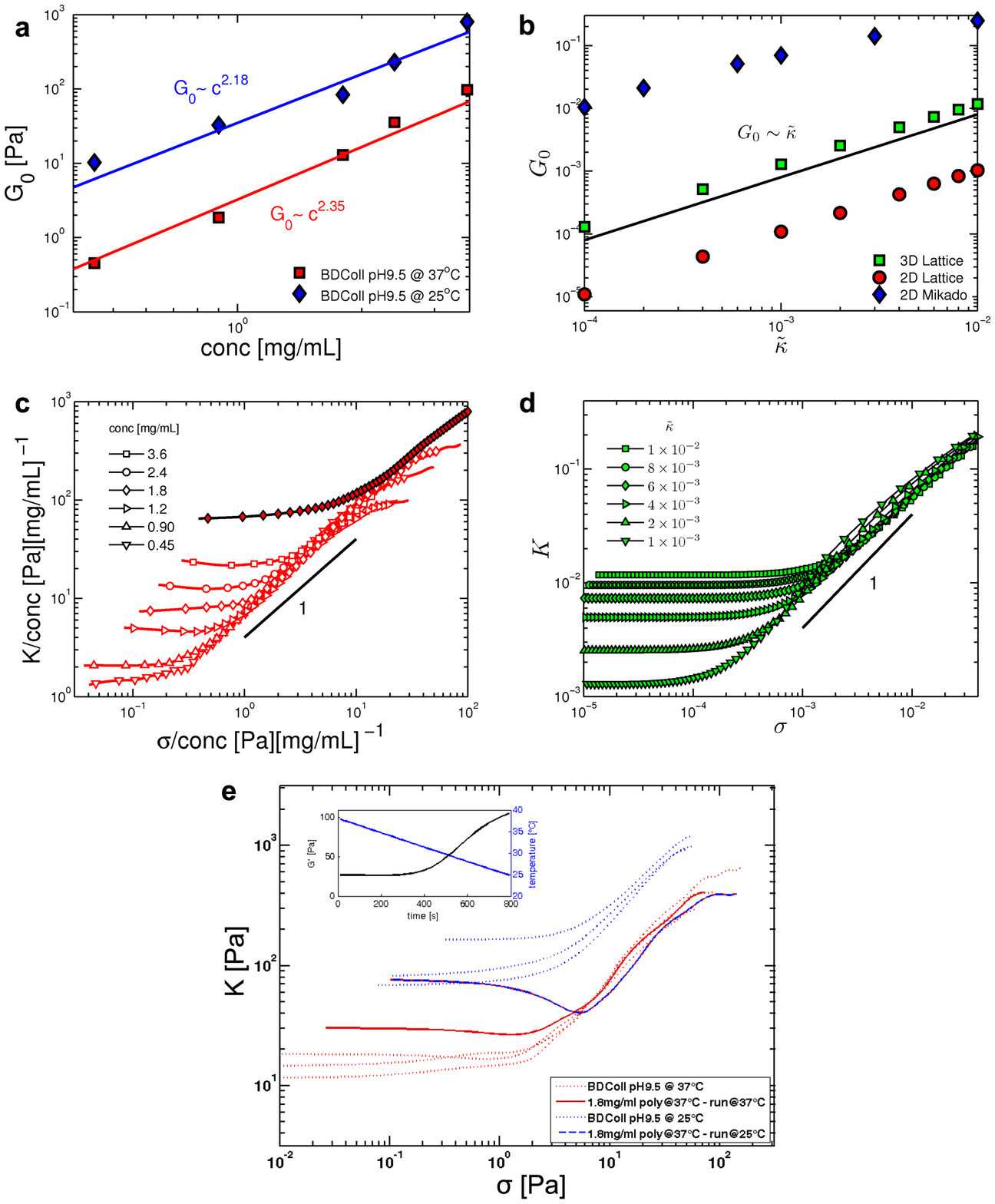}}
\caption{{\bf Concentration dependence of the linear shear modulus and stiffness vs stress curves.} (Color online) (a) Linear shear modulus vs protein concentration at different polymerization temperatures $37\,^{\circ}\mathrm{C}$ and $25\,^{\circ}\mathrm{C}$. (b) Linear modulus obtained from simulations on networks with different geometries: 2D/3D lattice and 2D Mikado. (c) Stiffness vs stress curves normalized by the concentration of collagen networks polymerized at $37\,^{\circ}\mathrm{C}$. For the network at a concentration of 1.8 mg/mL, 0.2\% glutaraldehyde (GA) cross-linkers are added (filled symbols) to increase the bending rigidity of the fibers. (d) Dimensionless stiffening curves from simulations on a 3D lattice for various fiber rigidities. (e) Stiffness vs stress curves showing the effect of running the rheology at $25\,^{\circ}\mathrm{C}$ for a network polymerized at $37\,^{\circ}\mathrm{C}$ (solid blue trace). For comparison, we show the result when the rheology is run at the same temperature as the polymerization at $37\,^{\circ}\mathrm{C}$ (solid red trace). The inset shows the increase in linear modulus (black trace) of a network polymerized at $37\,^{\circ}\mathrm{C}$ as the temperature cools down to $25\,^{\circ}\mathrm{C}$ with time (blue trace).}
\label{fig:conc}
\end{figure*}

\begin{figure*}[ht]
\centering
\centerline{\includegraphics[width=0.9\textwidth]{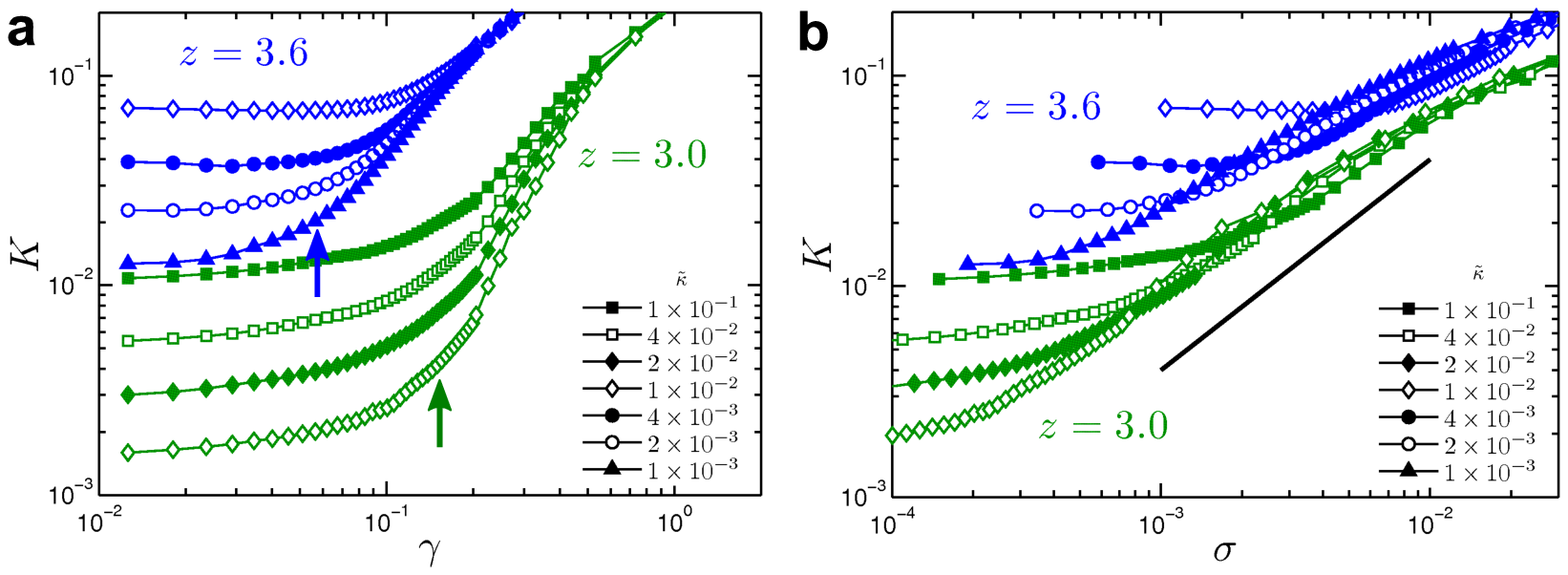}}
\caption{{\bf Shift of stiffening onset with network geometry.} (Color online) Shear stiffening of a 2D phantom network with average connectivity $z=3.6$ (blue) and $z=3.0$ (green). The aspect ratios are $L/\ell_0=5.2$ and $L/\ell_0=2.5$, respectively. (a). Stiffness versus strain shows the shift of the onset of stiffening to a lower value with increasing $z$ or $L/\ell_0$ indicated by the arrows. (b) Stiffness versus stress shows an increase in the amplitude ratio $K/\sigma$ as indicated by the upward shift of the curves in the stiffening regime with increasing $z$ or $L/\ell_0$. The solid line of unit slope serves as a visual guide.}
\label{fig:zdependence}
\end{figure*}
\begin{figure*}[hb!]
\centering
\centerline{\includegraphics[width=0.5\textwidth]{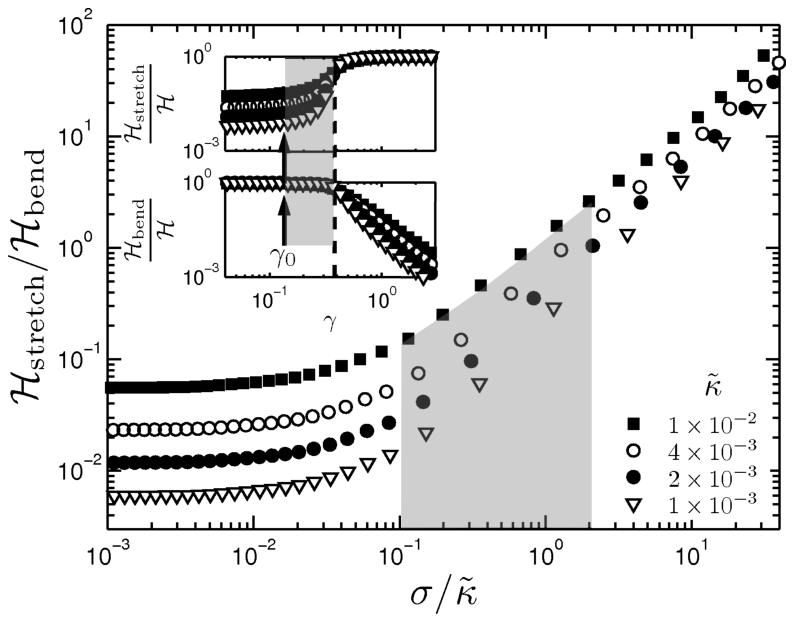}}
\caption{{\bf Stretching and bending contributions to the total energy.} The ratio of stretching energy to bending energy is less than unity in the stiffening regime, i.e., the shaded region from the critical strain $\gamma_0$ to the strain indicated by the thick dashed line. This shows that stretching modes are subdominant to bending in this regime. The insets show the relative contributions of stretching and bending to the total elastic energy.}
\label{fig:energyratio}
\end{figure*}

\begin{figure*}[ht]
\centering
\centerline{\includegraphics[width=0.5\textwidth]{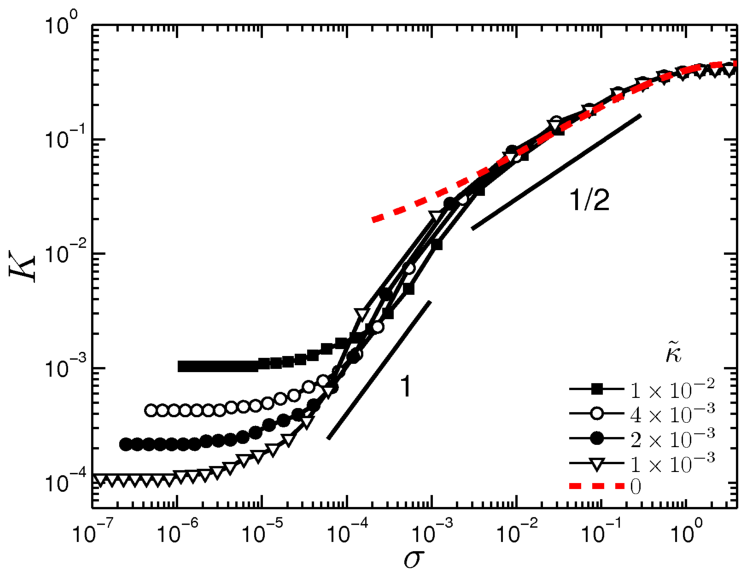}}
\caption{{\bf Stretch-dominated stiffening.} (Color online) Network simulation showing shear stiffening curves for various bending rigidities including the zero limit (red dashed curve). This limit corresponds to a network governed purely by stretching modes and as the figure shows, it leads to a different stiffening behavior where the modulus scales as $\sigma^{1/2}$. The small deviation from the slope of $1/2$ at low stress for the $\tilde{\kappa}=0$ limit is due to a finite-size effect. The line of unit slope only serves as guide to the eye.}
\label{fig:rope}
\end{figure*}
\begin{figure*}[hb]
\centering
\centerline{\includegraphics[width=0.5\textwidth]{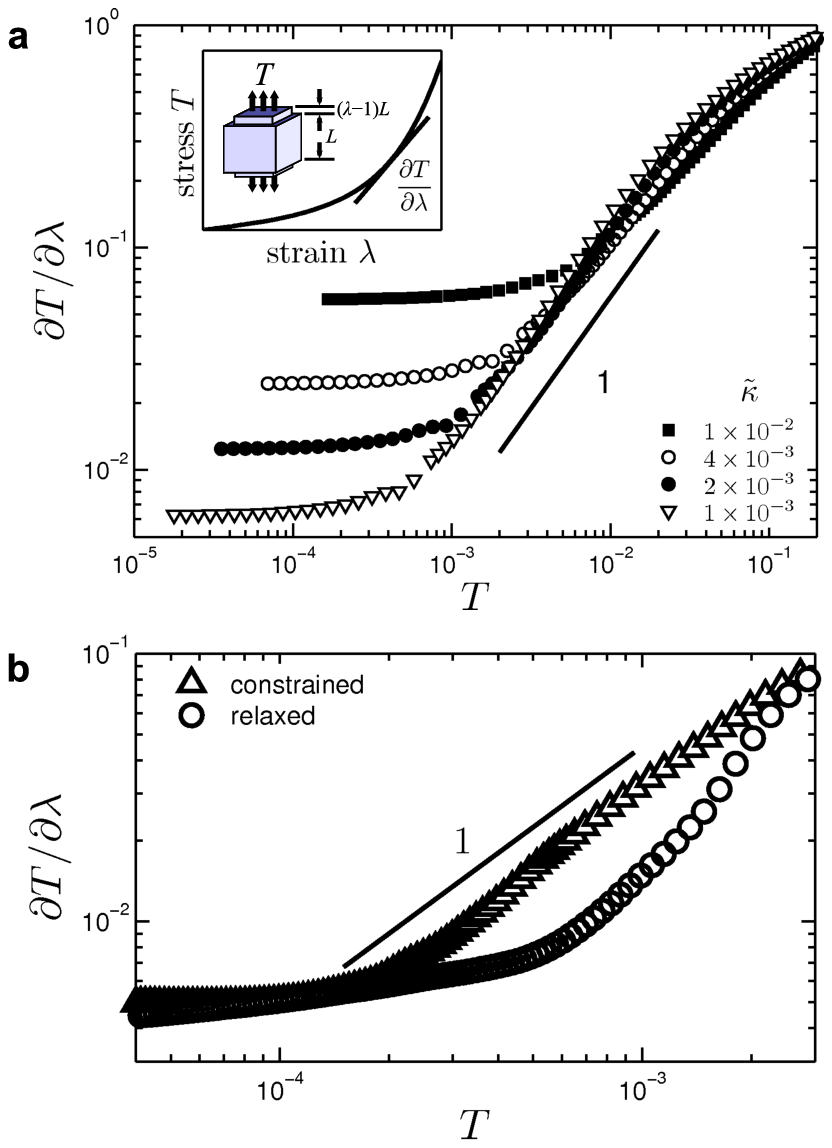}}
\caption{{\bf Stiffening under simple extension.} (Color online) Network simulation of stiffening under volume-preserving extension. (a) Stiffness vs tensile stress curve for a 3D network with different fiber rigidities. The line of unit slope is shown as a guide to the eye. The inset shows a schematic tensile stress vs tensile strain curve and how the stiffness is obtained. (b) Stiffening curve on a 2D network under extension with and without volume constraint.}
\label{fig:extension}
\end{figure*}

\end{document}